\documentclass[12pt]{article}

\usepackage{a4}

\usepackage{amsmath}

\usepackage{bm}
\usepackage{amsfonts}

\newcommand{\CC}{\mathbb{C}}
\newcommand{\ZZ}{\mathbb{Z}}

\newcommand{\tr}{{\rm tr}}
\newcommand{\wh}{\widehat}
\newcommand{\wt}{\widetilde}

\def\Pexp{\mathop{\rm Pexp}\nolimits}

\begin{document}\begin{titlepage}
\title{\vspace{-2cm}
\begin{flushright}
\normalsize{TIT/HEP-632\\
SNUTP13-005\\
November 2013}
\end{flushright}
       \vspace{2cm}
Factorization of ${\bm S}^3/\ZZ_n$ partition function
       \vspace{2cm}}
\author{
Yosuke Imamura\thanks{E-mail: \tt imamura@phys.titech.ac.jp} $^a$,\quad
Hiroki Matsuno\thanks{E-mail: \tt matsuno@th.phys.titech.ac.jp} $^a$,\quad
and\ \ Daisuke Yokoyama\thanks{E-mail: \tt ddyokoyama@snu.ac.kr} $^b$
\\[30pt]
{\it $^a$Department of Physics, Tokyo Institute of Technology,}\\
{\it Tokyo 152-8551, Japan}
\\
{\it $^b$Center for Theoretical Physics, Seoul National University,}\\
{\it Seoul 151-747, Korea}\\
}
\date{}

\maketitle
\thispagestyle{empty}

\vspace{0cm}

\begin{abstract}
\normalsize
We investigate $\bm S^3/\ZZ_n$ partition function of 3d $\mathcal N = 2$
supersymmetric field theories.
In a gauge theory the partition function is the sum of
the contributions of sectors specified by holonomies,
and we should carefully choose
the relative signs among the contributions.
We argue that the factorization to holomorphic blocks
is a useful criterion to determine the signs
and propose a formula for them.
We show that
the orbifold partition function of
a general non-gauge theory is correctly factorized
provided that we take appropriate relative signs.
We also present a few examples of gauge theories.
We point out that the sign factor for the orbifold partition function
is closely related to a similar sign factor in
the lens space index and
the 3d index.
\end{abstract}

\end{titlepage}

\tableofcontents

\section{Introduction}
In this paper, we discuss a rather technical issue concerning
the sign of the partition function of 3d ${\cal N}=2$ supersymmetric theories.
It is
defined by the path integral
\begin{equation}
Z_{\cal M}=\int{\cal D}\Phi e^{-S},
\label{zdef}
\end{equation}
where ${\cal M}$ is the background manifold and $\Phi$
collectively represents all dynamical fields in the theory.
The overall factor
is often ignored because it does not affect some observables
such as correlation functions.
However, in recent progress of supersymmetric field theories,
the partition function itself plays an important role.
For example, we can determine the superconformal R-charge of a 3d ${\cal N}=2$ theory at infra-red
fixed point by maximizing the real part of the free energy
$F=-\log Z_{{\bm S}^3}$
\cite{Jafferis:2010un,Jafferis:2011zi}.

In order to compute the partition function
including the overall factor unambiguously
through the path integral (\ref{zdef})
we need to fix the measure of the path integral carefully.
A convenient way to do this
is to exploit the fact that $Z_{\cal M}$ is obtained from
the supersymmetric index
of a 4d ${\cal N}=1$ theory as the small radius limit.
Let us consider the case of ${\bm S}^3$ partition function.
To obtain the 3d theory, we start from a 4d ${\cal N}=1$
theory in the background ${\bm S}^3\times{\bm S}^1$.
We define ${\bm S}^3$ by
\begin{equation}
|z_1|^2+|z_2|^2=1,\quad
z_1,z_2\in\CC.
\label{s3inc2}
\end{equation}
If we regard the ${\bm S}^1$ direction as a time,
the path integral of the 4d theory
is interpreted as the index
\begin{equation}
I(p_1,p_2,z_a)=\tr\left[
(-1)^Fq^{D-\frac{R}{2}}
(q^{-1}p_1)^{J_{1}+\frac{R}{2}}
(q^{-1}p_2)^{J_{2}+\frac{R}{2}}
z_a^{F_a}
\right],
\label{4dindex}
\end{equation}
where $F$, $R$, $D$, and $F_a$ are the fermion number,
the $U(1)_R$ charge, the dilatation, and
flavor charges, respectively.
$J_1$ and $J_2$ are the angular momenta
rotating $z_1$ and $z_2$, respectively.
The exponent of $q$ in (\ref{4dindex})
\begin{equation}
D-J_1-J_2-\frac{3}{2}R=\{Q,Q^\dagger\}
\end{equation}
is exact with respect to a supercharge $Q$,
and (\ref{4dindex}) is independent of the variable $q$.
Unlike the partition function $Z_{{\bm S}^3}$, there is a natural normalization
of $I$;
in the trace
over the Hilbert space,
every gauge invariant state contributes to the index by weight $1$,
and there is no ambiguity of the normalization
except for the signature.
The ${\bm S}^3$ partition function is obtained
as the $\beta\rightarrow 0$ limit of the index.
In non-supersymmetric theories the small radius limit
may in general diverge.
However, in the reduction from a 4d ${\cal N}=1$ theory to a 3d ${\cal N}=2$
theory that we consider here, we can obtain a finite result
due to the cancellation between bosonic
and fermionic contributions.\footnote
{
For this cancellation we should carefully include the zero-point contribution,
which is often neglected.
}
Therefore, once
we obtain the 4d index, we can unambiguously obtain the partition function
by the small radius limit
\cite{Dolan:2011rp,Gadde:2011ia,Imamura:2011uw}.
\begin{equation}
Z_{{\bm S}^3}(b,\mu_a)=
\lim_{\beta\rightarrow 0}I(p_i=e^{-\beta\omega_i},z_a=e^{-\beta\mu_a}),\quad
b=\sqrt{\frac{\omega_1}{\omega_2}}.
\label{smallradius}
\end{equation}
$\omega_i$ and $\mu_a$ are interpreted in the 3d theory as
squashing parameters and real mass parameters.
The partition function depends on $\omega_i$ through the single parameter
$b$
\cite{Imamura:2011uw,Hama:2011ea,Imamura:2011wg,Alday:2013lba,Nian:2013qwa,Closset:2013vra,Tanaka:2013dca}.

The ambiguity in the signature is due to the ambiguity in the statistics of the
vacuum state.
The statistics of states in the Hilbert space is fixed
once that of the vacuum state is specified.
However, there is no general rule to fix it,
and we need an additional criterion to
fix the overall sign.

For some use of $Z_{\cal M}$, like $F$ maximization, we only need the absolute value of $Z_{\cal M}$,
and one may think that the sign ambiguity is not important.
However, if the theory has multiple sectors,
we should sum up their contributions,
and we need to fix the relative signs among them.
This is the case when we consider a gauge theory
on a manifold
with non-trivial fundamental group.
In such a case
there are degenerate vacua labeled by
holonomies associated with non-trivial cycles.
In this paper we focus on the orbifold ${\bm S}^3/\ZZ_n$
defined from ${\bm S}^3$ in (\ref{s3inc2}) by the identification
\begin{equation}
(z_1,z_2)\sim (\omega z_1,\omega^{-1}z_2)\quad
\omega=e^{\frac{2\pi i}{n}}.
\label{zndef}
\end{equation}
The fundamental group of this manifold is $\ZZ_n$,
and vacua are labeled by $\ZZ_n$-valued holonomies
$h_a^{\rm (dyn)}$ associated with dynamical
$U(1)_a$ gauge symmetries
as well as continuous moduli parameters $\mu_a^{({\rm dyn})}$.
It is also possible to introduce
mass parameters $\mu _a^{\rm (ext)}$ and
non-trivial holonomies
$h_a^{\rm (ext)}$ for global $U(1)_a$ symmetries.
(We label both gauge and global symmetries by $a$.)
The partition function
is obtained by summing up the
contribution from sectors with
different $h_a^{\rm (dyn)}$
\begin{equation}
Z_{{\bm S}^3/\ZZ_n}(b,\mu_a^{(\rm ext)},h_a^{(\rm ext)})
=\sum_{h_a^{(\rm dyn)}}f(h)\int d\mu^{({\rm dyn})}
e^{-S_{{\bm S}^3/\ZZ_n}^{\rm cl}}Z^{\rm 1-loop}_{{\bm S}^3/\ZZ_n}(b,\mu_a,h_a),
\label{zorb}
\end{equation}
where $f(h)$ is the sign factor that we would like to determine,
and the explicit form of the integrand will be given in the next section.

A formula for the orbifold partition function has actually
been already given in \cite{Benini:2011nc}.
They derive the formula in two ways.
One is the orbifold projection from
${\bm S}^3$ partition function, and the other is
reduction from the lens space index, which is obtained
from ${\bm S}^3\times{\bm S}^1$ index by the orbifold projection.
In both derivations, they do not take account of the
possible emergence of non-trivial sign factors.
The formula has been used for some applications, and works well.
However, in some cases, we need to introduce extra sign factors.
For example, it is demonstrated in \cite{Imamura:2012rq}
for a few examples of dual pairs that
the matching of  the orbifold partition function of dual theories
requires non-trivial sign factors.

A similar problem of relative weight also arises in the
instanton sum in the ${\bm S}^4$ partition function.
It would be instructive to understand
how we can determine the relative weights
in that case,
before we explain our strategy
for the ${\bm S}^3/\ZZ_n$ partition function.

Let us consider an ${\cal N}=2$ supersymmetric gauge theory on ${\bm S}^4$.
By equivariant localization, we can localize the dynamics
of the theory
at two poles of ${\bm S}^4$,
and the partition function is written as \cite{Pestun:2007rz,Gomis:2011pf}
\begin{equation}
Z=\sum_{k_N,k_S=0}^{\infty}f(k_N,k_S)Z_N(k_N)Z_S(k_S),
\end{equation}
where $k_N$ and $k_S$ are, respectively,
the instanton number at the north pole and
the anti-instanton number at the south pole.
(Precisely, we need to perform the integral over the Coulomb branch parameterized
by constant scalar fields. Here we focus only on the instanton sum, and consider the contribution from a specific point in the Coulomb branch.)
We introduced the unknown phase factor $f(k_N,k_S)$.
This phase factor is strongly restricted by
assuming the locality of the theory \cite{Callan:1976je}.
If we assume the locality,
the path integral for the localized modes at the two poles should be
performed independently,
and thus the partition function is factorized into contributions
from the poles;
\begin{equation}
Z=
\sum_{k_N=0}^{\infty}
g(k_N)Z_N(k_N)
\sum_{k_S=0}^{\infty}
h(k_S)Z_S(k_S),
\end{equation}
where $g(k_N)$ and $h(k_S)$ are unknown phase factors.
Now we use the fact that disconnected components of
the configuration space of the theory is labeled
only by the total instanton number $k_N-k_S$.
This means that two configurations labeled by $(k^{(1)}_N,k^{(1)}_S)$ and
$(k^{(2)}_N,k^{(2)}_S)$ are in the same component of the configuration space
if $k_N^{(1)}-k_S^{(1)}=k_N^{(2)}-k_S^{(2)}$.
We can interpolate them by
continuous deformation and the relative phase between the contributions from
them can be in principle determined unambiguously by the continuity of
the action functional.
Here let us assume
for simplicity that there are no relative phases.
Namely, $g(k_N+n)h(k_S+n)$ does not depend on $n$.
Then, the relation
\begin{equation}
\frac{g(k_N+1)}{g(k_N)}
=\frac{h(k_S)}{h(k_S+1)}
\label{gh}
\end{equation}
holds.
The left-(right-)hand side of this equation is independent of $k_S$ ($k_N)$,
and (\ref{gh}) is a constant independent of both $k_N$ and $k_S$.
Let $c$ be the constant.
We obtain
\begin{equation}
g(k_N)=g(0)c^{k_N},\quad
h(k_S)=h(0)c^{-k_S},
\end{equation}
and the total partition function is
\begin{equation}
Z=f(0,0)\sum_{k_N,k_S=0}^{\infty}c^{k_N-k_S}Z_N(k_N)Z_S(k_S).
\end{equation}
Now we have determined the phase factor except the overall phase $f(0,0)$
and the constant $c$.
The factor $c^{k_N-k_S}$ can be identified with the
contribution of the topological $\theta$ term.

In this way, the relative phases for instanton sectors
of a 4d gauge theory can be fixed up to few parameters
by the factorization
of the partition function.
We take the same strategy to determine the relative
signs among holonomy sectors
of the orbifold partition function.
Actually, it is known that the ${\bm S}^3$ and ${\bm S}^2\times{\bm S}^1$
partition functions are factorized into
factors so-called holomorphic blocks \cite{Krattenthaler:2011da,Pasquetti:2011fj,Dimofte:2011py,Beem:2012mb,Taki:2013opa},
and a similar factorization is expected for the orbifold.
Each block is identified with the vortex partition function on a solid torus \cite{Pasquetti:2011fj,Beem:2012mb}.
In the following, we determine the relative signs of
holonomy sectors by requiring the factorization of the
orbifold partition function.

\section{Orbifold partition function}
\subsection{Naive projection}
Let us first summarize
how the formula for the
orbifold partition function $Z_{{\bm S}^3/\ZZ_n}$
is obtained
from the ${\bm S}^3$ partition function by naive $\ZZ_n$ orbifold projection \cite{Benini:2011nc}.

We consider a theory with general gauge group $G$ and matter representation $R$.
At a generic point in the Coulomb branch the gauge group $G$ is broken to its
Cartan subgroup $H$.
Let $V_a$, $W_\alpha$, and $\Phi_i$ denote the
vector multiplets for the Cartan part,
W-bosons, and chiral multiplets, respectively.
For later convenience,
we include external vector multiplets for global $U(1)_a$ symmetries
in $V_a$.
For distinction, we denote dynamical and non-dynamical components
by $V_a^{({\rm dyn})}$ and $V_a^{({\rm ext})}$, respectively.
The scalar components $\mu_a^{({\rm dyn})}$ for the
dynamical vector multiplets
parameterize the Coulomb branch
while those for external vector multiplets, $\mu_a^{({\rm ext})}$,
are real mass parameters.

By localization, we can reduce the path integral of an
${\cal N}=2$ supersymmetric theory on ${\bm S}^3$
into a finite dimensional matrix integral with
the integrand consisting of the classical and one-loop
factors:
\begin{equation}
Z_{{\bm S}^3}(\mu^{({\rm ext})})=\int d\mu^{({\rm dyn})} e^{-S^{\rm cl}_{{\bm S}^3}(\mu)}Z_{{\bm S}^3}^{\rm 1-loop}(\mu).
\label{zs3}
\end{equation}
The integration measure is given by
\begin{equation}
\int d\mu^{({\rm dyn})}\equiv \frac{1}{|W|}\prod_a\int_{-\infty}^\infty d\mu^{({\rm dyn})}_a
\end{equation}
where $a$ runs over dynamical part of $V_a$,
and $|W|$ is the order of the Weyl group of the gauge group.
The one-loop factor is the product of
the contributions of $W_\alpha$ and $\Phi_i$
\begin{equation}
Z_{{\bm S}^3}^{\rm 1-loop}=\frac{1}{\prod_\alpha s_b(\mu_\alpha+\frac{iQ}{2})}
\frac{1}{\prod_is_b(\mu_i-\frac{iQ}{2}(1-\Delta_i))},
\label{eq15}
\end{equation}
where $Q=b+b^{-1}$.
$\Delta_i$ is the Weyl weight of the scalar component of a chiral multiplet $\Phi_i$.
$\mu_\alpha$ and $\mu_i$ are the scalar components of the vector multiplets
coupling to $W_\alpha$ and $\Phi_i$;
\begin{equation}
\mu_\alpha=q_{\alpha a}\mu_a,\quad
\mu_i=q_{ia}\mu_a,
\label{muamui}
\end{equation}
where $q_{\alpha a}$ and $q_{ia}$ are the $U(1)_a$ charge of W-boson $W_\alpha$
and the chiral multiplet $\Phi_i$, respectively.
It is convenient to include R-charge in the charge matrix.
We define $q_{\alpha 0}$ and $q_{i0}$ as the R-charges
of the fermions in the multiplets $W_\alpha$ and $\Phi_i$,
\begin{equation}
q_{\alpha 0}=1,\quad
q_{i0}=\Delta_i-1,
\end{equation}
and we set the corresponding scalar parameter by
\begin{equation}
\mu_0\equiv\mu_R=\frac{iQ}{2}.
\end{equation}
Including the contribution of the $U(1)_R$ symmetry, we define
\begin{equation}
\wh\mu_\alpha=\mu_\alpha+\frac{iQ}{2},\quad
\wh\mu_i=\mu_i-\frac{iQ}{2}(1-\Delta_i).
\end{equation}
Then the one-loop factor (\ref{eq15}) is simply rewritten as
\begin{equation}
Z_{{\bm S}^3}^{\rm 1-loop}=\frac{1}{\prod_I s_b(\wh\mu_I)},
\end{equation}
where $I$ runs over both W-bosons and chiral multiplets.
$s_b(z)$ is the double sine function,
and can be expressed as an infinite product corresponding to the
spherical harmonic expansion on ${\bm S}^3$.
The contribution of a multiplet $I$
is
\begin{equation}
\frac{1}{s_b(\wh\mu_I)}
=\prod_{p,q=0}^\infty
\frac{b(q+\frac{1}{2})+b^{-1}(p+\frac{1}{2})+i\wh\mu_I}
{b(p+\frac{1}{2})+b^{-1}(q+\frac{1}{2})-i\wh\mu_I}.
\label{s3oneloop}
\end{equation}
The denominator and the numerator come from the bosonic and the fermionic
modes with angular momenta $(J_1,J_2)=(p,q)$, respectively.

The classical factor exists when the theory has
Chern-Simons terms.
If the action contains the Chern-Simons term
\begin{equation}
\frac{ik_{ab}}{4\pi}\int A_adA_b,
\label{eq22}
\end{equation}
the scalar field quadratic terms in the supersymmetric completion of
(\ref{eq22})
give the classical factor
\begin{equation}
e^{-S_{{\bm S}^3}^{\rm cl}(\mu)}=e^{-\pi i k_{ab}\mu_a\mu_b}.
\label{cscontribution}
\end{equation}

Let us move on to the orbifold partition function
$Z_{{\bm S}^3/\ZZ_n}$.
The orbifold ${\bm S}^3/\ZZ_n$ is defined from the
${\bm S}^3$ in
(\ref{s3inc2})
by the identification (\ref{zndef}).
The vacua
are parameterized by the scalar field $\mu_a$ and
holonomies
\begin{equation}
h_a=\frac{n}{2\pi}\oint _\gamma A_a,
\label{holodef}
\end{equation}
where $\gamma$ is a non-trivial loop in ${\bm S}^3/\ZZ_n$
generating the fundamental group.
$h_a$ is quantized to be integer,
and $h_a$ and $h_a+n$ are identified because they are
transformed to each other by a large gauge transformation.
We define $h_I=\{h_\alpha,h_i\}$, holonomies coupling to
$W_\alpha$ and $\Phi_i$, in a similar way to (\ref{muamui}):
\begin{equation}
h_I=\sum_a q_{Ia}h_a.
\label{holonomyI}
\end{equation}
We can turn on non-trivial holonomies for global symmetries
as well as the dynamical gauge symmetries.
Note that we will not turn on the holonomy for the R-symmetry
because it breaks supersymmetry.
(For a more general lens space $L(p,q)$, we need to turn on non-trivial
$U(1)_R$ holonomy to preserve supersymmetry.)

After the orbifold projection,
only modes compatible with the identification
(\ref{zndef}) contribute to the one-loop factor.
The condition for $\ZZ_n$ invariance
for modes of multiplet $I$
is
\begin{equation}
p-q=h_I\mod n,
\label{pqhn}
\end{equation}
and the one-loop factor is obtained by restricting the
product over $p$ and $q$ in (\ref{s3oneloop})
by the condition
(\ref{pqhn}).
We define the function
$s_{b,h_I}$ to express the contribution of each multiplet by
\begin{equation}
\frac{1}{s_{b,h_I}\left(\wh\mu_I\right)}
=\prod_{(p,q)\in\Lambda_{[h_I]}}
\frac{b(q+\frac{1}{2})+b^{-1}(p+\frac{1}{2})+i\wh\mu_I}{b(p+\frac{1}{2})+b^{-1}(q+\frac{1}{2})-i\wh\mu_I},
\label{sbh1}
\end{equation}
where
$\Lambda_{[h]}$ is
the set consisting of $(p,q)$ satisfying
the condition (\ref{pqhn}):
\begin{equation}
\Lambda_{[h]}
=\{(p,q)|p,q\geq0,p-q=h\mod n\}.
\end{equation}
The one-loop factor
(\ref{s3oneloop})
is replaced by
\begin{equation}
Z_{{\bm S}^3/\ZZ_n}^{\rm 1-loop}
=\prod_I\frac{1}{s_{b,h_I}\left(\wh\mu_I\right)}.
\end{equation}

For the classical part,
the $\ZZ_n$ orbifolding gives rise to
the extra $1/n$ factor in the action,
and
the Chern-Simons term gives the holonomy dependent phase \cite{Gang:2009wy,0209403,Griguolo:2006kp}:
\footnote{We find a slightly different formula
$e^{\frac{\pi i}{n}k_{ab}h_ah_b}$ for the holonomy dependent phase in the literature.
This is, however, not gauge invariant even for integer Chern-Simons levels.
A simple derivation of the holonomy dependent phase in (\ref{cscont}) is given in Appendix \ref{app:CSterm}.
}
\begin{equation}
e^{-S_{{\bm S}^3/\ZZ_n}^{\rm cl}(\mu,h)}
=e^{-\frac{\pi i}{n}k_{ab}\mu_a\mu_b}
 e^{-\frac{\pi i k_{ab}}{n}(n-1)h_ah_b}.
\label{cscont}
\end{equation}

The integration measure is
\begin{equation}
\int d\mu^{({\rm dyn})}
=\frac{1}{|W|}\prod_a\int_{-\infty}^\infty\frac{d\mu_a^{({\rm dyn})}}{n}.
\end{equation}
This is normalized
by using the relation to the lens space index \cite{Benini:2011nc}.

Combining these factors, we obtain the formula (\ref{zorb}).

Before ending this section, we would like to comment
on a subtlety lurking in the formula
(\ref{zorb}).
For the gauge invariance
of the factor (\ref{cscont}),
the Chern-Simons levels $k_{ab}$ must be integers.
This is, however, not always the case.
If the theory has parity anomaly,
some components of the bare Chern-Simons level should be
half odd integers to cancel the anomaly.
In this case, the factor (\ref{cscont})
itself is not gauge invariant.
Namely, it may change its sign under
a large gauge transformation that shifts $h_a$ by
$h_a\rightarrow h_a+nc_a$ ($c_a\in\ZZ$).
Of course, this is not an essential problem.
For the consistency,
we only need the
gauge invariance of the
whole integrand in
(\ref{zorb})
including the one-loop contribution.
In the following, we propose a general formula
for the sign factor $f(h)$, which will give the
sign of the partition function for
each holonomy sector in a gauge invariant way.

\subsection{Projection operator}
\label{subsec:pairing}
As we mentioned in the introduction, we use the factorization
to holomorphic blocks to determine the sign factor.
For ${\cal M}={\bm S}^3$ and ${\bm S}^2\times {\bm S}^1$,
it is known that the partition function is written in terms of
holomorphic blocks by \cite{Krattenthaler:2011da,Pasquetti:2011fj,Dimofte:2011py,Beem:2012mb,Taki:2013opa}
\begin{equation}
Z_{\cal M}=\sum_A B^A(x_a,q)B^A(\wt x_a,\wt q).
\label{factorization}
\end{equation}
In the case of ${\bm S}^3$,
the variables
$x_a$, $q$, $\wt x_a$, and $\wt q$ are given by
\begin{equation}
q=e^{2\pi i b^2},\quad
x_a=e^{2\pi b\mu_a},\quad
\wt q=e^{2\pi i b^{-2}},\quad
\wt x_a=e^{2\pi b^{-1}\mu_a}.
\label{s3arguments}
\end{equation}
This factorization is also expected for the orbifold partition function
with different definition for the variables
$x_a$, $q$, $\wt x_a$, and $\wt q$.
This factorization is naturally interpreted in Higgs branch localization,
in which the
index $A$ labels Higgs vacua and the blocks are identified with the vortex partition functions\cite{Pasquetti:2011fj,Beem:2012mb}.

In order to obtain a factorized form of the orbifold partition function,
it is convenient to rewrite $s_{b,h_I}$ in (\ref{sbh1}) as
\begin{equation}
\frac{1}{s_{b,h_I}\left(\wh\mu_I\right)}
=
{\cal P}_{(\wh\mu_I,h_I)}
\frac{1}{s_b(\wh\mu_I)}
\label{ps}
\end{equation}
where ${\cal P}_{(z,h)}$ with $z\in\CC$ and $h\in\ZZ_n$ is the operator acting on a function
of $z$ defined by
\begin{equation}
{\cal P}_{(z,h)} f(z)
=\prod_{(k,l)\in L_{h}} f(z_{k,l}),\quad
z_{k,l}\equiv \frac{z+ib(k-\frac{n-1}{2})+ib^{-1}(l-\frac{n-1}{2})}{n},
\label{Pdef}
\end{equation}
where
\begin{equation}
L_{h}=\{(k,l)|
0\leq k,l<n,\ 
k-l=h\mod n\}.
\end{equation}
An advantage of rewriting (\ref{sbh1})
with this operator is
that the operator preserves the factorized form of the function.
Namely, if a function $f(z)$ is the product of two functions
$g(z)$ and $h(z)$,
the relation ${\cal P}_{(z,h)}f(z)=({\cal P}_{(z,h)}g(z))({\cal P}_{(z,h)}h(z))$
holds.
Therefore, if the ${\bm S}^3$ partition function $Z_{{\bm S}^3}(\wh\mu)$
of a theory is factorized into holomorphic blocks as in (\ref{factorization}),
and if the orbifold partition function is obtained from $Z_{{\bm S}^3}(\wh\mu)$
by applying the operator ${\cal P}_{(\wh\mu,h)}$,
we can immediately obtain the factorized form of the
orbifold partition function by applying ${\cal P}_{(\wh\mu,h)}$
to the holomorphic blocks for ${\bm S}^3$.

The operator ${\cal P}_{(z,h)}$ is defined
to simplify the expression of
the $\ZZ_n$ projection of the one-loop factor,
and it is a priori not guaranteed that it correctly reproduces the
classical factor $e^{-S^{\rm cl}_{{\bm S}^3/\ZZ_n}}$ in the orbifold partition function.
Interestingly, up to the sign factor which
we have not fixed yet,
it reproduces the classical factor (\ref{cscont})
in the orbifold partition function
from (\ref{cscontribution}) for ${\bm S}^3$.
Let us consider a Chern-Simons term with factorized Chern-Simons level
$k_{ab}=\kappa c_ac_b$.
\begin{equation}
\frac{i\kappa c_ac_b}{4\pi}\int A_adA_b.
\label{cacb}
\end{equation}
For ${\bm S}^3$, this gives the classical factor
\begin{equation}
e^{-S^{\rm cl}_{{\bm S}^3}}=e^{-\kappa\pi i\wh\mu^2},\quad
\wh\mu=c_a\mu_a.
\end{equation}
Applying the operator ${\cal P}_{(\wh\mu,h)}$ on this function,
we obtain
\begin{align}
{\cal P}_{(\wh\mu,h)}e^{-\kappa\pi i\wh\mu^2}
&=
e^{\frac{\pi i\kappa}{n}\frac{(b+b^{-1})^2}{12}(n^2-1)}
\exp\left[
-\frac{\pi i\kappa}{n}
\left(\wh\mu^2
+[h](n-[h])\right)
\right],
\label{csxxxx}
\end{align}
where $h=c_ah_a$ and
$[h]$ denotes the smallest non-negative integer in
$h+n\ZZ$.
(We have not yet assumed that $\kappa$ is an integer.)
We rewrite this as
\begin{equation}
{\cal P}_{(\wh\mu,h)}e^{-\kappa\pi i\wh\mu^2}
=(\mbox{$b$-dependent factor})
\rho(h)^{2\kappa}
\exp\left[-\frac{\pi i\kappa}{n}\left(\wh\mu^2
 +(n-1)h^2\right)
 \right].
\label{eq40}
\end{equation}
In this paper, the factorization is used simply as a criterion
for the correct choice of sign factors, and we are not
interested in the prefactor depending only on $b$.
The exponential factor is nothing but the
classical factor in
(\ref{cscont}), and
$\rho(h)$ is
\begin{equation}
\rho(h)=
e^{\frac{\pi i}{2n}[h](n-[h])}
e^{-\frac{\pi i}{2n}(n-1)h^2}.
\label{rhoh}
\end{equation}
This function
always takes $+1$ or $-1$ depending on $h$.
As we will show shortly,
we can compose the sign factor $f(h)$ by using $\rho(h)$.

\subsection{Factorization and sign factor}
Let us consider a general non-gauge theory on ${\bm S}^3$.
As is pointed out in \cite{Beem:2012mb},
it is convenient to decompose the
Chern-Simons level into the part canceling the parity anomaly
and the remaining part.
\begin{equation}
k_{ab}=\sum_\alpha\kappa_\alpha c_{\alpha a} c_{\alpha b}
-\frac{1}{2}\sum_Iq_{Ia}q_{Ib}.
\label{decomposition}
\end{equation}
$\kappa_\alpha$ and $c_{\alpha a}$ in the first term are
integers, and the second term is the fractional contribution
that cancels the parity anomaly.
With this decomposition, we rewrite the partition function as
\begin{equation}
Z_{{\bm S}^3}(\mu)=e^{-\pi i k_{ab}\mu_a\mu_b}\frac{1}{\prod_Is_b(\wh\mu_I)}
=\prod_\alpha
(e^{-\pi i \mu_\alpha^2})^{\kappa_\alpha}
\prod_I
Z_\Delta(\wh\mu_I)
\label{froduct}
\end{equation}
where
$Z_\Delta(\wh\mu)$ is the partition function
of the ``tetrahedron theory,''\cite{Dimofte:2011ju} and given by
\begin{equation}
Z_\Delta(\wh\mu)=\frac{e^{\frac{\pi i}{2}\wh\mu^2}}{s_b(\wh\mu)}.
\label{tetra}
\end{equation}
The two kinds of factors in the product
(\ref{froduct})
are known to be factorized into holomorphic blocks \cite{Beem:2012mb}
\begin{align}
Z_\Delta(\mu)
&
=
e^{-\pi i\frac{b^2+b^{-2}}{24}}
B_\Delta(x,q)B_\Delta(\wt x,\wt q),\label{zdelta}\\
e^{-\pi i \mu^2}
&=
e^{\pi i\frac{b^2+b^{-2}}{12}}
B_{\rm CS}(x;q)
B_{\rm CS}(\wt x;\wt q).\label{zcs}
\end{align}
The blocks are given by
\begin{equation}
B_\Delta(x;q)=(qx^{-1};q),\quad
B_{\rm CS}(x;q)=\frac{1}{(-q^{\frac{1}{2}}x;q) (-q^{\frac{1}{2}}x^{-1};q)},
\label{holocs}
\end{equation}
where $(x;q)$ is the q-Pochhammer symbol
\begin{equation}
(x;q)=\prod _{k=0}^{\infty}(1-xq^k).
\end{equation}

An important feature of the factorization is that
the information of the background manifold
is encoded in the
definition of the arguments of holomorphic blocks.
For ${\bm S}^3$ they are given by (\ref{s3arguments}),
and the blocks for the orbifold should be given by the same functions
with different arguments.
We can confirm this by computing the holomorphic blocks for the orbifold by
applying the operator ${\cal P}_{(z,h)}$ to the
holomorphic blocks for ${\bm S}^3$.
Indeed, we can easily show
\begin{eqnarray}
&&
{\cal P}_{(\mu,h)}B_\Delta(x;q)=B_\Delta(x';q'),\quad
{\cal P}_{(\mu,h)}B_\Delta(\wt x;\wt q)=B_\Delta(\wt x';\wt q'),
\nonumber\\
&&
{\cal P}_{(\mu,h)}
B_{\rm CS}(x;q)=B_{\rm CS}(x',q'),\quad
{\cal P}_{(\mu,h)}
B_{\rm CS}(\wt x;\wt q)
=B_{\rm CS}(\wt x',\wt q'),
\end{eqnarray}
where the variables for the orbifold are
\begin{equation}
q'=\omega q^{\frac{1}{n}},\quad
x'=\omega^h x^{\frac{1}{n}},\quad
\wt q'=\omega\wt q^{\frac{1}{n}},\quad
\wt x'=\omega^{-h}\wt x^{\frac{1}{n}}.
\label{orbpairing}
\end{equation}


The arguments above guarantee that we obtain
the orbifold partition function that is correctly factorized into the
holomorphic blocks by simply applying the projection operator
to the factors in ${\bm S}^3$ partition function
(\ref{froduct}).
\begin{align}
Z_{{\bm S}^3/\ZZ_n}(\mu,h)
&\propto
\prod_\alpha
({\cal P}_{(\mu_\alpha,h_\alpha)}e^{-\pi i \mu_\alpha^2})^{\kappa_\alpha}
\prod_I
{\cal P}_{(\wh\mu_I,h_I)}Z_\Delta(\wh\mu_I)
\nonumber\\
&\propto
\prod_\alpha
e^{\frac{\pi i}{n}\kappa_\alpha(\wh\mu_\alpha^2+(n-1)h_\alpha^2)}
\prod_I
\frac{\rho(h_I)e^{\frac{\pi i}{2n}(\wh\mu_I^2+(n-1)h_I^2)}}{s_{b,h_I}(\wh\mu_I)}
\nonumber\\
&=
e^{-S^{\rm cl}_{{\bm S}^3/\ZZ_n}(\mu,h)}
\prod_I
\frac{\rho(h_I)}{s_{b,h_I}(\wh\mu_I)}
\end{align}
``$\propto$'' means the ignorance of prefactors depending only on $b$.
By comparing this to (\ref{zorb}),
we obtain
\begin{equation}
f(h)=\prod_I\rho(h_I).
\end{equation}
(We cannot fix the overall sign factor independent of $h$, which
we are not interested in.)
We can absorb the sign factor by the redefinition of the orbifold
double sine function
\begin{equation}
s_{b,h}^{\rm imp}(z)=\rho(h)s_{b,h}(z),
\label{improved}
\end{equation}
and then we can present the
orbifold partition function in the same form with the
original one.
\begin{equation}
Z_{{\bm S}^3/\ZZ_n}(\mu,h)=e^{-S_{{\bm S}^3/\ZZ_n}^{\rm cl}(\mu,h)}\frac{1}{\prod_Is_{b,h_I}^{\rm imp}(\wh\mu_I)}.
\label{improvedz}
\end{equation}

In \cite{Imamura:2012rq}, a similar improvement of the orbifold
double sine function
$\wh s_{b,h}(z)=\sigma_h s_{b,h}(z)$
is proposed for odd $n=2m+1$.
The extra sign factor $\sigma_h$ is related to $\rho(h)$ through
\begin{equation}
\sigma_h=(-)^{mh}\rho(h).
\end{equation}
In \cite{Imamura:2012rq},
only theories without parity anomaly are considered.
This means that the charge assignment $q_{Ia}$ for every $U(1)$
gauge symmetry satisfies
$\sum_I q_{Ia}\in 2\ZZ$.
In such a case,
the difference of $\sigma_h$ and $\rho(h)$ does not affect the
partition function.
In the case of even $n$,
\cite{Imamura:2012rq} 
did not succeed in finding such an improvement.
The reason is that
in \cite{Imamura:2012rq}
the sign factor $\sigma_h$ is assumed to be a periodic function of $h$ with period $n$.
The function $\rho(h)$ does not satisfy this condition.
It may change its sign under the shift $h\rightarrow h+n$.
\begin{equation}
\frac{\rho(h+n)}{\rho(h)}=(-1)^{(n-1)h+\frac{n(n-1)}{2}}.
\end{equation}
This means that the improved function $s_{b,h}^{\rm imp}$ may
change its sign in the large gauge transformation
$h\rightarrow h+n$.
This, however, does not cause any problem.
If the parity anomaly arising in the one-loop part is
correctly canceled by the bare Chern-Simons term,
the partition function
(\ref{improvedz}) is invariant under the shift $h\rightarrow h+n$.

\subsection{Gauge theories}
In the previous subsection we gave a prescription to fix the relative signs
among holonomy sectors.
It is implemented by the redefinition of the orbifold double sine function
(\ref{improved}).
The orbifold partition function of an arbitrary non-gauge theory with this improvement
is correctly factorized into holomorphic blocks.

We expect that this improvement works for gauge theories, too.
Namely, the orbifold partition function
\begin{equation}
Z_{{\bm S}^3/\ZZ_n}(\mu^{({\rm ext})},h^{({\rm ext})})=\sum_{h^{({\rm dyn})}}\int d\mu^{({\rm dyn})} e^{-S_{{\bm S}^3/\ZZ_n}^{\rm cl}(\mu,h)}\frac{1}{\prod_Is_{b,h}^{\rm imp}(\wh\mu_I)}
\label{improvedzgauge}
\end{equation}
for a gauge theory is correctly factorized into holomorphic blocks
as in (\ref{factorization}).
Unfortunately, we have not succeeded in proving this for an arbitrary gauge theory.
We here consider two examples of gauge theories,
SQED with $N_f=1$, and an $su(2)$ Chern-Simons theory with an adjoint chiral multiplet.

\subsubsection{SQED}
As the first example, let us consider SQED with one flavor $(q,\wt q)$.
This theory has four $U(1)$ symmetries.
One is a gauge symmetry $U(1)_G$, and the others are global symmetries.
See Table \ref{table:1} for charge assignments.
\begin{table}[htb]
\caption{The charge assignments in the SQED and XYZ model.
For $U(1)_R$ symmetry the charges of the fermion components are shown.}
\label{table:1}
\begin{center}
\begin{tabular}{cccccccc}
\hline
\hline
&& $q$ & $\wt q$ & & $X$ & $Y$ & $Z$ \\
\hline
$U(1)_R$ && $\Delta-1$ & $\Delta-1$ & & $-\Delta$ & $-\Delta$ & $2\Delta-1$ \\
$U(1)_G$ && $1$ & $-1$ & & - & - & - \\
$U(1)_V$ && $0$ & $0$ & & $1$ & $-1$ & $0$ \\
$U(1)_A$ && $1$ & $1$  && $-1$ & $-1$ & $2$ \\
\hline
\end{tabular}
\end{center}
\end{table}
$U(1)_R$ is an R-symmetry.
$U(1)_A$ is a flavor symmetry acting on $q$ and $\wt q$ with charge $+1$.
$U(1)_V$ is the topological symmetry,
and the corresponding external gauge field $A_V$
couples to the $U(1)_G$ flux $dA_G$ through the Chern-Simons term
\begin{equation}
\frac{1}{2\pi}\int A_VdA_G.
\end{equation}
The partition function is
\begin{equation}
Z^{{\bm S}^3/\ZZ_n}_{\rm SQED}
=\sum_{h_G=0}^{n-1}\int \frac{e^{\frac{2\pi i}{n}\mu_V\mu_G}e^{\frac{2\pi i}{n}h_Vh_G}}
{s_{b,h_q}^{\rm imp}(\wh\mu_q)s_{b,h_{\wt q}}^{\rm imp}(\wh
\mu_{\wt q})}\frac{d\mu_G}{n},
\label{zaqed}
\end{equation}
where $\wh\mu_I$ and $h_I$ are defined by
\begin{equation}
\wh\mu_q=(\Delta-1)\frac{iQ}{2}+\mu_A+\mu_G,\quad
\wh\mu_{\wt q}=(\Delta-1)\frac{iQ}{2}+\mu_A-\mu_G.
\end{equation}
\begin{equation}
h_q=h_A+h_G,\quad
h_{\wt q}=h_A-h_G.
\end{equation}

Let us confirm that this orbifold partition function is factorized
into holomorphic blocks.
Although it would not be difficult to directly prove the factorization by
performing the integral by using the residue theorem,
we take another way.
In \cite{Imamura:2012rq}, it is confirmed that
the orbifold partition function of the SQED with an appropriate choice of the
sign factor coincides with that of the XYZ model, the system consisting of three chiral multiplets
$X$, $Y$, and $Z$ interacting through the superpotential $W=XYZ$.
Because XYZ model is a non-gauge theory and we have already proved the
factorization of non-gauge theories, this duality relation guarantees the
factorization of the partition function of the SQED.

The charge assignments for XYZ model is also shown in Table \ref{table:1},
and the partition function is given by
\begin{equation}
Z^{{\bm S}^3/\ZZ_n}_{\rm XYZ}=\frac{1}{
s_{b,h_X}^{\rm imp}(\wh\mu_X)
s_{b,h_Y}^{\rm imp}(\wh\mu_Y)
s_{b,h_Z}^{\rm imp}(\wh\mu_Z)}
\label{zxtz}
\end{equation}
with the parameters
\begin{equation}
\wh\mu_X=-\Delta\frac{iQ}{2}-\mu_A+\mu_V,\quad
\wh\mu_Y=-\Delta\frac{iQ}{2}-\mu_A-\mu_V,\quad
\wh\mu_Z=(2\Delta-1)\frac{iQ}{2}+2\mu_A.
\end{equation}
and
\begin{equation}
h_X=-h_A+h_V,\quad
h_Y=-h_A-h_V,\quad
h_Z=2h_A.
\end{equation}
It is easy to numerically check the coincidence of
(\ref{zaqed}) and (\ref{zxtz}).
If we use the result of \cite{Imamura:2012rq},
what we have to do to confirm the relation
$Z_{\rm XYZ}=Z_{\rm SQED}$
is to show the sign factor determined in \cite{Imamura:2012rq}
by requiring $Z_{\rm XYZ}=Z_{\rm SQED}$ is same as
the extra sign factor introduced by replacing $s_{b,h}$ by $s_{b,h}^{\rm imp}$.
Indeed,
the product of the five sign factors corresponding to the five
double sine functions in $Z_{\rm SQED}$ and $Z_{\rm XYZ}$
coincides with the factor given
in \cite{Imamura:2012rq}.
\begin{equation}
\sigma(h_G,h_V,h_A)=
\rho(h_q)
\rho(h_{\wt q})
\rho(h_X)
\rho(h_Y)
\rho(h_Z).
\end{equation}

\subsubsection{$su(2)$ gauge theory}
Next, as a simple example of non-Abelian gauge theory,
we consider an $su(2)_1$ gauge theory
coupled to an adjoint chiral multiplet $\Phi$.
(We use $su(2)$ instead of $SU(2)$ to emphasize we do not specify
the global structure of the gauge group.)
Jafferis and Yin \cite{Jafferis:2011ns} proposed that
this theory is dual to the theory 
consisting a single chiral multiplet $X\sim {\rm tr}\Phi ^2$.
If we can show the matching of the partition functions
for the dual pair
the factorization of the $su(2)$ theory is guaranteed
as the previous example.

In general, we cannot completely specify a gauge theory only
by local information.
There may be different theories distinguished by
global structure of the gauge group.
In \cite{Aharony:2013hda}
importance of such distinction
in four-dimensional supersymmetric gauge theories
is pointed out,
and it is investigated
how such theories are related to each other by Seiberg duality.
The duality is checked in \cite{Razamat:2013opa}
by matching the lens space index,
which is sensitive to the global structure of the gauge group.
Similar aspects in three-dimensional gauge theories are
studied in \cite{Aharony:2013kma}.

The $su(2)$ theory we discuss here
is also an example of such a theory.
It has only an adjoint chiral multiplet as a matter field,
and no elementary fields are transformed by the center of $SU(2)$.
Therefore, precisely speaking,
there are two choices of the gauge group,
$SU(2)$ and $SO(3)$.
Although it is important problem to clarify
how the different choices of the gauge group
affect the duality and the factorization,
we will not do any detailed analysis here.
We only present the result of a preliminary analysis
based on the numerical computation of the partition function.

The symmetries and the charge assignments
for the dual theories are shown in Table \ref{symJY}.
\begin{table}[htb]
\caption{The symmetries and the charge assignments in Jafferis-Yin duality.
The $U(1)_R$ charges in the table are those for fermion component of the multiplets.}
\label{symJY}
\begin{center}
\begin{tabular}{ccccc}
\hline
\hline
&& $\Phi$ & & $X$ \\
\hline
$SU(2)_G$ && {\rm adj} && - \\
$U(1)_R$ && $\Delta-1$ & & $2\Delta-1$ \\
$U(1)_A$ && $1$ && $2$ \\
\hline
\end{tabular}
\end{center}
\end{table}
In the $su(2)$ theory,
the action contains the Chern-Simons term
\begin{equation}
\frac{i}{4\pi}\int \tr_{\rm fund}\left(A_GdA_G-\frac{2i}{3}A_G^3\right)
\label{su2csterm}
\end{equation}
for the dynamical gauge field, and
\begin{align}
\frac{(-3/2)i}{4\pi}\int ( (\Delta -1)A_R+A_A) d( (\Delta -1)A_R+A_A)+\frac{i}{4\pi}\int A_RdA_R
\end{align}
for the external $U(1)_R$ and $U(1)_A$ gauge fields.
On the other hand,
in the chiral free theory contains the Chern-Simons term
\begin{align}
\frac{(-1/2)i}{4\pi}\int ((2\Delta -1)A_R+2A_A)d((2\Delta -1)A_R+2A_A)
\end{align}
for the external gauge fields.

We define the holonomy parameter $h$ for the dynamical $su(2)$ gauge group
by
\begin{equation}
U=\exp\left(i\oint_\gamma A\right) =
\begin{pmatrix}
e^{\frac{\pi i h}{n}} \\
& e^{-\frac{\pi i h}{n}}
\end{pmatrix}.
\end{equation}
If the gauge group is $SU(2)$,
$U^n$ must be the unit matrix, and the holonomy is quantized by
$h\in 2\ZZ$, while for $SO(3)$ gauge group $h$ can be an
arbitrary integer.
The periodicity of $h$ also depends on the global structure of the gauge group
and the background manifold,
and there are two possibilities, $h\sim h+2n$ or $h\sim h+n$.
Here we will not argue which of these possibilities
should be adopted.
We simply compute the partition functions for all the holonomy sectors
labeled by $h=0,1,\ldots,2n-1$, and
infer from the duality which sectors should be summed up.

The orbifold partition function of each holonomy sector of the $su(2)$ theory is
\begin{align}
Z^{{\bm S}^3/\ZZ_n}_{su(2)}(h,h_A)&=\int \frac{d\mu _G}{2n}e^{-\frac{\pi i}{2n}\mu _G^2}e^{-\frac{\pi i}{2n}(n-1)h^2}e^{\frac{\pi i}{n}\left[ \frac{3}{2}\left\{ (\Delta -1)\frac{iQ}{2}+\mu _A \right\} ^2+\frac{Q^2}{4} \right]}e^{\frac{3\pi i}{2n}(n-1)h_A^2}\nonumber \\
&\quad \times \frac{1}{s^{\rm imp}_{b,h_{\phi _+}}(\wh \mu _{\phi _+}) s^{\rm imp}_{b,h_{\phi _0}}(\wh \mu _{\phi _0}) s^{\rm imp}_{b,h_{\phi _-}}(\wh \mu _{\phi _-}) s^{\rm imp}_{b,h_{W^+}}(\wh \mu _{W^+}) s^{\rm imp}_{b,h_{W^-}}(\wh \mu _{W^-})},
\end{align}
where the parameters are given by
\begin{gather}
\wh \mu _{\phi _+}
=(\Delta -1)\frac{iQ}{2}+\mu _A+\mu _G,\quad \wh \mu _{\phi _0}
=(\Delta -1)\frac{iQ}{2}+\mu _A,\quad \wh \mu _{\phi _-}
=(\Delta -1)\frac{iQ}{2}+\mu _A-\mu _G,\nonumber\\
\wh \mu _{W^+}=\frac{iQ}{2}+\mu _G,\quad \wh \mu _{W^-}=\frac{iQ}{2}-\mu _G,\\
h_{\phi _+}=h_A+h,\quad h_{\phi _0}=h_A,\quad h_{\phi _-}=h_A-h,\quad h_{W^+}=h,\quad h_{W^-}=-h.
\end{gather}
The orbifold partition function of the chiral free theory is
\begin{align}
Z^{{\bm S}^3/\ZZ_n}_X(h_A)=e^{\frac{\pi i}{2n}\left\{ (2\Delta -1)\frac{iQ}{2}+2\mu _A\right\} ^2}e^{\frac{2\pi i}{n}(n-1)h_A^2}\frac{1}{s^{\rm imp}_{b,h_X}(\wh \mu _X)},
\end{align}
where the parameters are given by
\begin{align}
\wh \mu _X=(2\Delta -1)\frac{iQ}{2}+2\mu _A,\quad h_X=2h_A.
\end{align}

Numerical results are divided to four cases:
\begin{itemize}
\item In the case of $n\in 4{\mathbb Z}$,
only the even sector coincides up to a constant factor.
\begin{align}
\sum _{h=0,2,\ldots ,2n-2}Z^{{\bm S}^3/\ZZ_n}_{su(2)}(h,h_A)=2Z^{{\bm S}^3/\ZZ_n}_X(h_A).
\end{align}
\item In the case of $n\in 4{\mathbb Z}+1$,
both even and odd sector coincides up to a constant factor.
\begin{align}
\sum_{h=0,2,\ldots ,2n-2} Z^{{\bm S}^3/\ZZ_n}_{su(2)}(h,h_A)
=\sum_{h=1,3,\ldots ,2n-1} Z^{{\bm S}^3/\ZZ_n}_{su(2)}(h,h_A)
=\sqrt2e^{\frac{\pi i}{4}}Z^{{\bm S}^3/\ZZ_n}_X(h_A).
\end{align}
\item In the case of $n\in 4{\mathbb Z}+2$,
only the odd sector coincides up to a constant factor.
\begin{align}
\sum_{h=1,3,\ldots ,2n-1}Z^{{\bm S}^3/\ZZ_n}_{su(2)}(h,h_A)=-2iZ^{{\bm S}^3/\ZZ_n}_X(h_A).
\end{align}
\item In the case of $n\in 4{\mathbb Z}+3$,
both even and odd sector coincides with different constant
factor.
\begin{gather}
\sum _{h=0,2,\ldots ,2n-2}Z^{{\bm S}^3/\ZZ_n}_{su(2)}(h,h_A)=\sqrt2e^{\frac{\pi i}{4}}Z^{{\bm S}^3/\ZZ_n}_X(h_A),\\
\sum_{h=1,3,\ldots ,2n-1}Z^{{\bm S}^3/\ZZ_n}_{su(2)}(h,h_A)=-\sqrt2e^{\frac{\pi i}{4}}Z^{{\bm S}^3/\ZZ_n}_X(h_A).
\end{gather}
\end{itemize}
In all cases, we do not need additional sign factor
if we use the improved function $s_{b,h}^{\rm imp}$.
These results strongly suggests that
we should take the following holonomy sectors to sum up.
\begin{itemize}
\item $n\in 4\ZZ$: $h=0,2,\ldots,2n-2$.
\item $n\in 4\ZZ+1$: $h=0,2,\ldots,2n-2$ or $h=1,3,\ldots,2n-1$.
\item $n\in 4\ZZ+2$: $h=1,3,\ldots,2n-1$.
\item $n\in 4\ZZ+3$: $h=0,2,\ldots,2n-2$ or $h=1,3,\ldots,2n-1$.
\end{itemize}
Unfortunately,
we have no clear explanation for these non-trivial choices of the holonomy sectors.
We hope we can return to this problem in future work.

\section{The sign factor  to the lens space index and the 3d index}

In this section we discuss the effect of the sign factor $f(h)$
derived in Subsection \ref{subsec:pairing} to the 4d lens space index \cite{Benini:2011nc} and 3d superconformal index \cite{Kim:2009wb,Imamura:2011su,Kapustin:2011jm}.
Since the orbifold partition function is related to the lens space index as well as
the 3d superconformal index through dimensional reductions
the same or similar sign factor should appear in those indices.
Let us first summarize the lens space index and its reductions, and then, we discuss the sign factors.

The lens space index can be obtained by the orbifold projection of the 4d index on ${\bm S}^3 \times {\bm S}^1$.
The projection is performed along the Hopf fiber direction of the ${\bm S}^3$ of the ${\bm S}^3 \times {\bm S}^1$,
and it is realized by leaving the modes compatible with the identification (\ref{zndef}).
Since the rotation along the Hopf fiber direction is characterized by $(J_1 -J_2)$ the projection is realized by
inserting the operator
\begin{align}
 \frac{1}{n} \sum_{m=0}^{n-1} e^{2\pi i \frac{m}{n} (J_1-J_2)}
 \label{orbop}
\end{align}
to the index (\ref{4dindex}).
Namely, the lens space index $I_n$ is
\begin{align}
 I_n (p_1,p_2,z_a) = \frac{1}{n} \sum_{m=0}^{n-1} I(e^{\frac{2 \pi i}{n}m}p_1,e^{-\frac{2 \pi i}{n}m}p_2,e^{-\frac{2 \pi i}{n}h_a}z_a)  ,
\end{align}
where we introduced the holonomies for global symmetries $h_a$.

Let us focus on a theory consisting of a single chiral multiplet $\Phi$ whose charge for a $U(1)$ global symmetry is 1.
The 4d index on ${\bm S}^3\times {\bm S}^1$ defined by (\ref{4dindex}) for the theory can be rewritten as follows \cite{Romelsberger:2007ec}.
\begin{align}
 I_\Phi (p_1,p_2,z) &= \Pexp' \left[ I_\Phi^\mathrm{sp} \right] \\
 I_\Phi^\mathrm{sp} (p_1,p_2,z) &= \sum_{i,j=0}^\infty \left(z p_1^i p_2^j -z^{-1} p_1^{j+1} p_2^{i+1} \right)
\label{spphi}
 \\
 &= \frac{z}{1-p_1 p_2} \left(\frac{1}{1-p_1} +\frac{p_2}{1-p_2}\right)
 -\frac{p_1 p_2 z^{-1}}{1-p_1 p_2} \left(\frac{p_1}{1-p_1} +\frac{1}{1-p_2}\right)  ,
\end{align}
where $z$ is a fugacity for the global symmetry,
and the prime of the plethystic exponential denotes that it includes the zero point contributions; $\Pexp' [x] = e^{x/2} \Pexp[x]$.
In this form the insertion of the operator (\ref{orbop}) is equivalent to leave the modes invariant under the condition
\begin{align}
 i - j = h \mod n
 \label{condforind}
\end{align}
in (\ref{spphi}).
Then, the lens space index for this theory can be written as follows.
\begin{align}
 I_{n,\Phi} (z,h) &= \Pexp'[I_{n,\Phi}^\mathrm{sp}(z,h)] \\
 I_{n,\Phi}^\mathrm{sp}(z,h) &= \frac{z}{1-p_1 p_2} \left(\frac{p_1^{[h]}}{1-p_1^{n}} +\frac{p_2^{n-[h]}}{1-p_2^{n}}\right)
 -\frac{p_1 p_2 z^{-1}}{1-p_1 p_2} \left(\frac{p_1^{n-[h]}}{1-p_1^{n}} +\frac{p_2^{[h]}}{1-p_2^{n}}\right)  .
\end{align}

Let us now consider the effect of the sign factor (\ref{rhoh}).
The orbifold partition function is derived in the small radius limit (\ref{smallradius})
of the lens space index.
Since the sign factor is independent of the radius $\beta$ it can be uplifted to the lens space index,
and
we define the improved index as
\begin{equation}
I_n^\mathrm{imp} (z,h)=\left( \prod_I\rho(h_I)\right) I_n (z,h)  .
\label{IandI0}
\end{equation}
For the theory considered above the improved lens space index is written as
\begin{align}
 I_{n,\Phi}^\mathrm{imp} (z,h) &= e^{-\frac{i \pi}{2} h(1-h)} \Pexp'[I_{n,\Phi}^\mathrm{imp, sp}(z,h)] \\
 I_{n,\Phi}^\mathrm{imp, sp} (z,h) &= \frac{z}{1-p_1 p_2} \left(\frac{p_1^{h}}{1-p_1^{n}} +\frac{p_2^{n-h}}{1-p_2^{n}}\right)
 -\frac{p_1 p_2 z^{-1}}{1-p_1 p_2} \left(\frac{p_1^{n-h}}{1-p_1^{n}} +\frac{p_2^{h}}{1-p_2^{n}}\right)  .
\end{align}
Note that $[h]$ in $I_{n,\Phi}$ is now replaced by $h$.
Although both $I_{n,\Phi}$ and $\rho(h)$ include $[h]$
the combination of those (\ref{IandI0}) is analytic in terms of $h$.

In order to illustrate the effect of the phase factor we obtained to the 3d index,
we first describe
the 3d index for the tetrahedron theory, which is written as follows
\begin{align}
 I^\mathrm{3d}_{\Delta} &= (-1)^{-\frac{1}{2}|h|} \left( (-q^{\frac{1}{2}})^{\frac{|h|-h}{2} } z^{-\frac{|h|-h}{2} } \right) \Pexp \left[ \frac{q^{\frac{|h|}{2} } z}{1-q}
 -\frac{q^{1+\frac{|h|}{2} } z^{-1}}{1-q} \right] .
\label{dimofte}
\end{align}
Later, it is noticed in \cite{Dimofte:2011py} that the 3d index can be written in an analytic form
by adding a simple phase factor:
\begin{align}
 I^\mathrm{3d, imp}_{\Delta} &= i^{|h| } I^\mathrm{3d}_{\Delta} =
 \Pexp \left[ \frac{q^{\frac{h}{2} } z}{1-q} -\frac{q^{1+\frac{h}{2} } z^{-1}}{1-q} \right]  .
\end{align}
The phase factor is needed for the 3d index to be factorized.
As is pointed out in \cite{Benini:2011nc},
the 3d index is obtained from the 4d lens space index by taking $n\rightarrow\infty$ limit.
When we take this limit $h$ is kept finite and is identified with the
magnetic flux in ${\bm S}^2\times {\bm S}^1$.
In this limit the sign factor (\ref{rhoh}) can be rewritten as
\begin{equation}
\rho(h)=
e^{\frac{\pi i}{2}(|h|-h^2)}  .
\label{rohlargeh}
\end{equation}
The contribution of the classical factor in (\ref{cscont}) in the large $n$ limit becomes
\begin{equation}
e^{-\pi ik_{ab}h_ah_b}  .
\end{equation}
As the tetrahedron theory has Chern-Simons term with level $-1/2$
its contribution combined with the extra sign factor (\ref{rohlargeh}) is
\begin{equation}
e^{\frac{\pi i}{2}h^2}\times
e^{\frac{\pi i}{2}(|h|-h^2)}
=
i^{|h|}  .
\label{3dphase}
\end{equation}
This is precisely the factor introduced in (\ref{dimofte})
to improve the 3d index.
Note that the improved 3d index is analytic in terms of $h$, which is inherited from the analytic structure
of the improved 4d index.

\section{Conclusions}
We focused on the sign of the orbifold partition function.
We proposed a formula that systematically determine the relative signs
among holonomy sectors.
We use the factorization to the holomorphic blocks as a criterion
to determine correct signs.
For non-gauge theories we proved that the partition function
with signs determined by the formula is correctly factorized into the
holomorphic blocks.
In the case of gauge theories,
as a simple example, we considered two theories:
SQED with $N_f=1$ and $su(2)$ gauge theory with an adjoint flavor.
For the former we confirmed the factorization of the orbifold partition
function.
In the case of the $su(2)$ theory, we have not fully understood
the summation over the holonomy sectors.
We guessed which holonomy sector contributes to the
partition function with the help of the duality to a non-gauge
theory, which is known as Jafferis-Yin duality.
We found that we have to choose an appropriate subset of the
holonomy sectors to obtain the partition function
consistent to the duality.
The formula for the sign factor
gives the correct relative phases for the contributing
sectors.
Therefore, up to the subtlety for the choice of the
holonomy sector, which is probably related to the
global aspects of the gauge bundle on the orbifold,
the formula seems to give appropriate signs.

We also discussed the sign factors in the lens space index of
the 4d theories, and the index for 3d theories.
The formulae for these indices have been known, and
contain non-trivial sign factors,
which are often implicit in the literature.
We showed that they are closely related to
the sign factor for the orbifold partition function.

\section*{Acknowledgments}
We would like to thank S.~Terashima for useful information and S.~Kim for valuable discussion.
Y.~I. is partially supported by Grant-in-Aid for Scientific Research
(C) (No.24540260), Ministry of Education, Science and Culture, Japan.
H.~M. acknowledges the financial support from the
Center of Excellence Program by MEXT, Japan through the
``Nanoscience and Quantum Physics'' Project of the Tokyo
Institute of Technology.
The work of D.Y. is supported in part by
the National Research Foundation of Korea Grant NRF-2012R1A2A2A02046739,
the Global Center of Excellence Program by MEXT
Japan through the ``Nanoscience and Quantum Physics'' Project of the Tokyo
Institute of Technology, and
Perimeter Institute for Theoretical Physics. Research at Perimeter Institute is supported by
the Government of Canada through Industry Canada and by
the Province of Ontario through the Ministry of Economic Development \& Innovation.

\appendix
\section{Appendix}
\subsection{Classical contribution of Chern-Simons terms}
\label{app:CSterm}
Let us compute
the Chern-Simons term
\begin{equation}
S_{\rm CS}
=\frac{ik}{4\pi}\int_{\cal M} AdA
=\frac{ik}{4\pi}\int_{X} F\wedge F,
\label{csterm}
\end{equation}
for a flat $U(1)$ connection.
$X$ is a manifold whose boundary is ${\cal M}={\bm S}^3/\ZZ_n$.
To compute $e^{-S_{\rm CS}}$ unambiguously for an
arbitrary integer $k$, $X$ should be a spin manifold \cite{Witten:2003ya}.

We represent ${\cal M}={\bm S}^3/\ZZ_n$ as
the Hopf fibration over $B={\bm S}^2$.
Let us represent $B$ as the boundary of the solid cylinder $C$:
$r\leq 1$, $0\leq z\leq n+1$,
where $(r,\phi,z)$ is the cylindrical coordinate system
in three-dimensional space.
We divide $\cal M$ into $n+1$ regions: $M_0,M_1,\ldots,M_n$.
The $\ell$-th region $M_\ell$ is defined by $\ell\leq z\leq \ell+1$.
Let $0\leq \psi_\ell<2\pi$ be the fiber coordinate in the region $M_\ell$.
On the boundary between adjacent regions, the fiber coordinates
are related by
\begin{equation}
\psi_\ell|_{z=\ell}=(\psi_{\ell+1}+\phi)|_{z=\ell}.
\end{equation}

We want to realize a flat connection
with the holonomy
\begin{equation}
h=\frac{n}{2\pi}\oint_\gamma A,
\end{equation}
where $\gamma$ is a cycle along an ${\bm S}^1$ fiber.
Let us consider the following gauge potential in $M_\ell$.
\begin{equation}
A_\ell|_{M_\ell}=\frac{h}{n}d\psi_\ell+c_\ell d\phi,
\label{abdr}
\end{equation}
For the gauge field on the top ($z=n+1$) and the bottom ($z=0$) of the cylinder
to be smooth and flat, we need to set
\begin{equation}
c_0=c_n=0.
\label{c0cn}
\end{equation}
The gauge field jumps on the boundaries by
\begin{equation}
A_\ell-A_{\ell-1}|_{z=\ell}=\left(c_\ell-c_{\ell-1}-\frac{h}{n}\right)d\phi.
\end{equation}
For this to be a gauge symmetry,
the coefficients $c_\ell$ should satisfy
\begin{equation}
c_\ell-c_{\ell-1}-\frac{h}{n}\in\ZZ.
\label{cdiff}
\end{equation}

To compute the Chern-Simons action for this gauge potential,
we need to define the manifold $X$ and to extend the gauge potential
into $X$.
We adopt $X$ which is topologically $n$-centered Taub-NUT.
We represent $X$ as an ${\bm S}^1$ fibration over $C$
with $n$ centers placed on the axis at $z=1,2,\ldots,n$.
$X$ is also divided into $n+1$ regions $X_0,X_1,\ldots,X_n$ at $z=1,2,\ldots, n$.
We take the following ansatz for the extension of the gauge field
in the $\ell$-th region.
\begin{equation}
A_\ell=\left[\frac{h}{n}-B(z)(1-f(r))\right]d\psi_\ell+c_\ell f(r)d\phi,
\label{ainside}
\end{equation}
where $B(z)$ and $f(r)$ are continuous functions satisfying
\begin{equation}
B(0)=B(n+1)=0,\quad
f(0)=0,\quad
f(1)=1.
\end{equation}
These conditions guarantee that (\ref{ainside})
coincides to (\ref{abdr}) on $\cal M$.

Let us think about the smoothness of the gauge field
at the center at $z=\ell$.
The center is located on the boundary between $M_\ell$ and $M_{\ell-1}$.
In each region the gauge field near the center is given by
\begin{equation}
A_\ell=b_\ell d\psi_\ell,\quad
A_{\ell-1}=b_\ell d\psi_{\ell-1},\quad
b_\ell=\frac{h}{n}-B(\ell).
\end{equation}
If we assume
\begin{equation}
b_\ell\in\ZZ,
\label{bkzz}
\end{equation}
these gauge field can be eliminated by the gauge transformation
\begin{equation}
A_\ell\rightarrow A_\ell'=A_\ell-b_\ell d\psi_\ell,\quad
A_\ell\rightarrow A_{\ell-1}'=A_{\ell-1}-b_\ell d\psi_{\ell-1},
\label{atoap}
\end{equation}
and then the gauge field is smooth at the center.
For the gauge potential after the gauge transformation
(\ref{atoap})
the jump of the gauge field on the boundary is
\begin{equation}
A'_\ell-A'_{\ell-1}|_{z=\ell}
=(c_\ell-c_{\ell-1}-B(\ell))f(r)d\phi.
\end{equation}
and $A_\ell'$ and $A'_{\ell-1}$ are smoothly connected
if
\begin{equation}
c_\ell-c_{\ell-1}=B(\ell)\quad
(1\leq \ell\leq n).
\end{equation}
Note that
the condition (\ref{bkzz}) follows
from this and
(\ref{cdiff}).

By using relations we obtained above,
we can easily compute the integral of the instanton density
over $X_\ell$.
\begin{eqnarray}
\int_{X_\ell} F_\ell\wedge F_\ell
&=&
8\pi^2c_\ell\left[B(z)\right]_\ell^{\ell+1}
\left[f(r)-\frac{1}{2}f(r)^2\right]_0^1
\nonumber\\
&=&
4\pi^2c_\ell(c_{\ell+1}-2c_\ell-c_{\ell-1}).
\label{intxk}
\end{eqnarray}
This depends on the constants $c_\ell$ that satisfy the conditions
(\ref{c0cn}) and
(\ref{cdiff}).
Although there are infinitely many solutions to these conditions
and the integral
(\ref{intxk}) depends on the choice of a solution,
the classical action is uniquely determined
up to the unphysical shift by $2\pi i\ZZ$.
\begin{align}
S=\frac{ik}{4\pi}\int_X F\wedge F
=\pi ik\sum_{\ell=1}^{n-1} c_\ell(c_{\ell+1}-2c_\ell+c_{\ell-1})
=\pi ik\frac{1-n}{n}h^2\mod 2\pi i.
\end{align}
We can easily extend this result to general Chern-Simons term
for multiple abelian gauge fields
\begin{align}
S=\frac{ik_{ab}}{4\pi}\int_{{\bm S}^3/\ZZ_n} A_adA_b
=\pi i\frac{1-n}{n}k_{ab}h_ah_b\mod 2\pi i.
\end{align}


\begin{thebibliography}{99}

\bibitem{Jafferis:2010un} 
  D.~L.~Jafferis,
  ``The Exact Superconformal R-Symmetry Extremizes Z,''
  JHEP {\bf 1205}, 159 (2012)
  [arXiv:1012.3210 [hep-th]].

\bibitem{Jafferis:2011zi} 
  D.~L.~Jafferis, I.~R.~Klebanov, S.~S.~Pufu and B.~R.~Safdi,
  ``Towards the F-Theorem: N=2 Field Theories on the Three-Sphere,''
  JHEP {\bf 1106}, 102 (2011)  [arXiv:1103.1181 [hep-th]].

\bibitem{Dolan:2011rp}
  F.~A.~H.~Dolan, V.~P.~Spiridonov, G.~S.~Vartanov,
  ``From 4d superconformal indices to 3d partition functions,''
  [arXiv:1104.1787 [hep-th]].
  
\bibitem{Gadde:2011ia}
  A.~Gadde and W.~Yan,
  ``Reducing the 4d Index to the $S^3$ Partition Function,''
  arXiv:1104.2592 [hep-th].

\bibitem{Imamura:2011uw}
  Y.~Imamura,
  ``Relation between the 4d superconformal index and the S$^3$ partition function,''
  [arXiv:1104.4482 [hep-th]].

\bibitem{Hama:2011ea} 
  N.~Hama, K.~Hosomichi and S.~Lee,
  ``SUSY Gauge Theories on Squashed Three-Spheres,''
  JHEP {\bf 1105}, 014 (2011)
  [arXiv:1102.4716 [hep-th]].
  
\bibitem{Imamura:2011wg} 
  Y.~Imamura and D.~Yokoyama,
  ``N=2 supersymmetric theories on squashed three-sphere,''
  Phys.\ Rev.\ D {\bf 85}, 025015 (2012)
  [arXiv:1109.4734 [hep-th]].

\bibitem{Alday:2013lba} 
  L.~F.~Alday, D.~Martelli, P.~Richmond and J.~Sparks,
  arXiv:1307.6848 [hep-th].  

\bibitem{Nian:2013qwa} 
  J.~Nian,
  ``Localization of Supersymmetric Chern-Simons-Matter Theory on a Squashed $S^3$ with $SU(2)\times U(1)$ Isometry,''
  arXiv:1309.3266 [hep-th].

\bibitem{Closset:2013vra} 
  C.~Closset, T.~T.~Dumitrescu, G.~Festuccia and Z.~Komargodski,
  ``The Geometry of Supersymmetric Partition Functions,''
  arXiv:1309.5876 [hep-th].

\bibitem{Tanaka:2013dca} 
  A.~Tanaka,
  ``Localization on round sphere revisited,''
  arXiv:1309.4992 [hep-th].

\bibitem{Benini:2011nc} 
  F.~Benini, T.~Nishioka and M.~Yamazaki,
  ``4d Index to 3d Index and 2d TQFT,''
  Phys.\ Rev.\ D {\bf 86}, 065015 (2012)
  [arXiv:1109.0283 [hep-th]].

\bibitem{Imamura:2012rq} 
  Y.~Imamura and D.~Yokoyama,
  ``$S^3/Z_n$ partition function and dualities,''
  JHEP {\bf 1211}, 122 (2012)
  [arXiv:1208.1404 [hep-th]].

\bibitem{Pestun:2007rz} 
  V.~Pestun,
  ``Localization of gauge theory on a four-sphere
  and supersymmetric Wilson loops,''
  Commun.\ Math.\ Phys.\  {\bf 313}, 71 (2012)  [arXiv:0712.2824 [hep-th]].

\bibitem{Gomis:2011pf} 
  J.~Gomis, T.~Okuda and V.~Pestun,
  ``Exact Results for 't Hooft Loops in Gauge Theories on $S^4$,''
  JHEP {\bf 1205}, 141 (2012)  [arXiv:1105.2568 [hep-th]].


\bibitem{Callan:1976je} 
  C.~G.~Callan, Jr., R.~F.~Dashen and D.~J.~Gross,
  Phys.\ Lett.\ B {\bf 63}, 334 (1976).

\bibitem{Krattenthaler:2011da} 
  C.~Krattenthaler, V.~P.~Spiridonov and G.~S.~Vartanov,
  ``Superconformal indices of three-dimensional theories related by mirror symmetry,''
  JHEP {\bf 1106}, 008 (2011)
  [arXiv:1103.4075 [hep-th]].

\bibitem{Pasquetti:2011fj} 
  S.~Pasquetti,
  ``Factorisation of N = 2 Theories on the Squashed 3-Sphere,''
  JHEP {\bf 1204}, 120 (2012)  [arXiv:1111.6905 [hep-th]].

\bibitem{Dimofte:2011py} 
  T.~Dimofte, D.~Gaiotto and S.~Gukov,
  ``3-Manifolds and 3d Indices,''
  arXiv:1112.5179 [hep-th].

\bibitem{Beem:2012mb} 
  C.~Beem, T.~Dimofte and S.~Pasquetti,
  ``Holomorphic Blocks in Three Dimensions,''
  arXiv:1211.1986 [hep-th].

\bibitem{Taki:2013opa} 
  M.~Taki,
  ``Holomorphic Blocks for 3d Non-abelian Partition Functions,''
  arXiv:1303.5915 [hep-th].

\bibitem{Gang:2009wy} 
  D.~Gang,
  ``Chern-Simons theory on L(p,q) lens spaces and Localization,''
  arXiv:0912.4664 [hep-th].
\bibitem{0209403}
    S. K. Hansen, T. Takata
    ``Reshetikhin-Turaev invariants of Seifert 3-manifolds
    for classical simple Lie algebras, and their asymptotic expansions,''
    arXiv:math/0209403 [math.GT]
\bibitem{Griguolo:2006kp} 
  L.~Griguolo, D.~Seminara, R.~J.~Szabo and A.~Tanzini,
  ``Black holes, instanton counting on toric singularities
  and q-deformed two-dimensional Yang-Mills theory,''
  Nucl.\ Phys.\ B {\bf 772}, 1 (2007)
  [hep-th/0610155].

\bibitem{Dimofte:2011ju} 
  T.~Dimofte, D.~Gaiotto and S.~Gukov,
  ``Gauge Theories Labelled by Three-Manifolds,''
  arXiv:1108.4389 [hep-th].

\bibitem{Jafferis:2011ns} 
  D.~Jafferis and X.~Yin,
  ``A Duality Appetizer,''
  arXiv:1103.5700 [hep-th].

\bibitem{Aharony:2013hda} 
  O.~Aharony, N.~Seiberg and Y.~Tachikawa,
  ``Reading between the lines of four-dimensional gauge theories,''
  JHEP {\bf 1308}, 115 (2013)
  [arXiv:1305.0318 [hep-th]].

\bibitem{Razamat:2013opa} 
  S.~S.~Razamat and B.~Willett,
  ``Global Properties of Supersymmetric Theories and the Lens Space,''
  arXiv:1307.4381 [hep-th].

\bibitem{Aharony:2013kma} 
  O.~Aharony, S.~S.~Razamat, N.~Seiberg and B.~Willett,
  ``3$d$ dualities from 4$d$ dualities for orthogonal groups,''
  JHEP {\bf 1308}, 099 (2013)
  [arXiv:1307.0511 [hep-th]].

\bibitem{Kim:2009wb} 
  S.~Kim,
  ``The Complete superconformal index for N=6 Chern-Simons theory,''
  Nucl.\ Phys.\ B {\bf 821}, 241 (2009)
  [Erratum-ibid.\ B {\bf 864}, 884 (2012)]
  [arXiv:0903.4172 [hep-th]].
\bibitem{Imamura:2011su} 
  Y.~Imamura and S.~Yokoyama,
  ``Index for three dimensional superconformal field theories with general R-charge assignments,''
  JHEP {\bf 1104}, 007 (2011)
  [arXiv:1101.0557 [hep-th]].
\bibitem{Kapustin:2011jm} 
  A.~Kapustin and B.~Willett,
  ``Generalized Superconformal Index for Three Dimensional Field Theories,''
  arXiv:1106.2484 [hep-th].

\bibitem{Romelsberger:2007ec}
  C.~Romelsberger,
  ``Calculating the Superconformal Index and Seiberg Duality,''
  arXiv:0707.3702 [hep-th].

\bibitem{Witten:2003ya} 
  E.~Witten,
  ``SL(2,Z) action on three-dimensional conformal field theories with Abelian symmetry,''
  In *Shifman, M. (ed.) et al.: From fields to strings, vol. 2* 1173-1200
  [arXiv:hep-th/0307041].


\end{thebibliography}
\end{document}